# Plasma-enhanced atomic layer deposition of $Al_2O_3$ on graphene using monolayer hBN as interfacial layer

*Bárbara Canto, Martin Otto, Michael J. Powell, Vitaliy Babenko, Aileen O'Mahony, Harm Knoops, Ravi S. Sundaram, Stephan Hofmann, Max C. Lemme and Daniel Neumaier**

Dr. Bárbara Canto, Prof. Dr. Daniel Neumaier
Chair of Smart Sensor Systems, Wuppertal University, Wuppertal 42119, Germany
E-mail: dneumaier@uni-wuppertal.de
Martin Otto, Prof. Dr. Max C. Lemme, Prof. Dr. Daniel Neumaier
AMO GmbH, Advanced Microelectronic Center Aachen, Aachen 52074, Germany
Dr. Michael J Powell, Dr. Aileen O'Mahony, Dr. Harm Knoops, Dr. Ravi S Sundaram
Oxford Instruments Plasma Technology UK, Bristol BS494AP,
United Kingdom
Prof. Dr. Max C. Lemme
Chair of Electronic Devices, RWTH Aachen University, Aachen
52074, Germany
Dr. Vitaliy Babenko, Prof. Dr. Stephan Hofmann
Department of Engineering, University of Cambridge, Cambridge
CB3 0FA, United Kingdom

The deposition of dielectric materials on graphene is one of the bottlenecks for unlocking the potential of graphene in electronic applications. In this paper we demonstrate the plasma enhanced atomic layer deposition of 10 nm thin high quality $Al_2O_3$ on graphene using a monolayer of hBN as protection layer. Raman spectroscopy was performed to analyze possible structural changes of the graphene lattice caused by the plasma deposition. The results show that a monolayer of hBN in combination with an optimized deposition process can effectively protect graphene from damage, while significant damage was observed without an hBN layer. Electrical characterization of double gated graphene field effect devices confirms that the graphene did not degrade during the plasma deposition of $Al_2O_3$. The leakage current densities were consistently below 1 pA/µm² for electric fields across the insulators of up to 8 MV/cm, with irreversible breakdown happening above. Such breakdown electric fields are typical for $Al_2O_3$ and can be seen as an indicator for high quality dielectric films.

## 1. Introduction

Graphene has remarkable properties that make this material interesting for science and applications.[1] Especially in electronics, graphene has attracted significant attention because of its excellent electronic and photonic properties including high carrier mobility and broadband optical response.[2-10] One of the essential steps in manufacturing graphene based electronic and photonic devices is to deposit a dielectric on top of this material, either for encapsulation or for electrical gating.[11,12] Specifically, electro-optical modulators for data communication require the deposition of thin dielectrics with high breakdown field and low leakage current on top of graphene, without reducing the quality of the graphene.[13-14] Substantial efforts have been put into the deposition of high-k dielectrics on graphene for different device applications. Atomic layer deposition (ALD) is considered the most appropriate technique, in particular for thin layers as it provides precision control of the uniformity, composition control and film thickness.[15-17] However, ALD growth requires the chemical adsorption of the precursors on the substrate surface, so the chemical inertness of graphene makes the growth of dielectrics by ALD difficult. There are different types of ALD processes for the deposition of oxides like $Al_2O_3$, where water, ozone or oxygen plasma is used as an oxygen source together with an Al precursor, typically trimethylaluminum (TMA). Direct growth of high-quality and thin dielectrics on graphene by a thermal, water-based ALD process is not possible without process adaptations due to the lack of available nucleation sites on graphene. For a total surface coverage and a high oxide uniformity the presence of seed layers on the graphene or other pretreatments that facilitate the initial growth of a closed layer of dielectric on its surface its required.[16, 18-23] A seed layer may be deposited prior to ALD to promote the growth of a closed film of dielectric, for example e-beam evaporation of Al before depositing $Al_2O_3$ by water-based ALD.[18,19] Another possibility is a modification of the deposition process during the ALD deposition.[16, 21-24] which has been achieved by increased water absorption at lower temperatures just before the actual water-based ALD process at higher temperatures[16, 21] or by deposition of a TMA-based functionalization layer either by $NO_2$-TMA deposition at room temperature,[22] by omitting the purging steps after TMA and $H_2O$ pulses for the first several cycles of ALD[23] or by increasing the residence time of the precursors in the deposition chamber.[24] However, these methods require additional steps in the fabrication and the majority of these processes also introduce defects in graphene that can be seen by Raman spectroscopy.[19, 21, 23]

The deposition of $Al_2O_3$ on graphene before transfer, i.e. on the substrate where graphene was grown is another alternative. Strong interaction between the growth substrate and graphene can enhance the nucleation of $Al_2O_3$ deposited by thermal ALD. In the case of monolayer graphene on native metallic substrates like Copper the presence of polar traps at the interface (graphene/metal) can increase the adsorption of water molecules onto graphene in the thermal process using water as precursor.[25-27] In the case of graphene grown on SiC substrates, the adsorption energy for water molecules can be enhanced by the high n-type doping of the graphene induced by the underlying buffer layer.[28,29] However, if the oxide/graphene stack needs to be transferred to a different substrate, the handling and bending of the film during transfer introduces strain in the graphene and damage to the oxide and graphene films. The bending can induce cracks that may lead to significant reduction of film quality.[26] Plasma-enhanced atomic layer deposition (PEALD), on the other hand, has significant advantages over the thermal-based process: shorter process times and higher quality oxides are possible, enabling thinner gate oxides with higher breakdown fields[30]. However, the oxygen plasma process can cause significant surface damage especially to sensitive materials like graphene. The interface between the two-dimensional crystal graphene and the three dimensional (3D) amorphous $Al_2O_3$ is also critical for the electrical properties of graphene.[31] Graphene's intrinsic exceptional electrical properties are best preserved when there are only van der Waals interactions with the surrounding dielectrics, while $Al_2O_3$ tends to form covalent bonds to the underlying layer. This leads to non-perfect interfaces, reducing the mobility in graphene, increasing 1/f noise and introducing reliability problems of graphene-based devices.[32] Single-crystal multilayer hBN has been shown to improve the electronic performance of graphene immensely, leading to higher mobilities at room temperature due to weak interlayer van der Waals forces in both graphene and hBN and the small lattice mismatch between hBN and graphite (less than 2% [31]). While scalable chemical vapor deposition (CVD) processes for hBN have not yet reached a maturity comparable to graphene, poly-crystalline hBN films are widely available, with mono-layer hBN being most well controlled in terms of domain size (mm size demonstrated) and homogeneity. Such CVD monolayer hBN has recently been used as interfacial layer on single crystalline graphene flakes to enable encapsulation with 17 nm SiN as dielectric deposited by using plasma-enhanced (CVD).[32] Recently, thin films of hBN (1.6 nm) have been used as a protective buffer layer to protect graphene from damage during PEALD of $Al_2O_3$.[33] For very thin dielectrics, where the higher quality of dielectrics deposited by PEALD becomes important, a 1.6 nm buffer layer may be too thick and despite the thickness of the hBN there was still some damage to the graphene. These works demonstrate an increased interest in being able to protect graphene from plasma damage.[33]

In this work we demonstrate that even a monolayer of hBN on-top of graphene can act as an effective protective and interfacial layer for the graphene when PEALD with a very short plasma duration is used, enabling PEALD deposition of high quality and 10 nm thin $Al_2O_3$. Similar to graphene, hBN also has no dangling bonds, which can prevent proper nucleation in a thermal ALD process.  However, also the hBN is expected to receive slight damage due to the plasma exposure, which can act as nucleation sites for the $Al_2O_3$ growth process. Therefore, no additional seeding layers or pretreatments are needed for the PEALD process on hBN. For this study, we have intentionally chosen monolayer hBN grown by CVD, because it is available in wafer-scale (15 cm) areas and, due to the thin body of atomic monolayer thickness, it only gives a small contribution to the total dielectric thickness.
All experiments were conducted with monolayer graphene and monolayer hBN grown by CVD on metal foils. The hBN/graphene stacks were prepared by sequential etching and transfer steps and transferred to a 6 inch Si wafer covered by 90 nm thermally grown $SiO_2$ (see Experimental section for details). In the reference samples only graphene was transferred without hBN. After the transfer, the samples were encapsulated by plasma-assisted atomic layer deposition (PEALD) with 10 nm of $Al_2O_3$ in an Oxford Instruments Atomfab™ ALD system using TMA and oxygen plasma as precursors. It is important to point out that the plasma time was only 0.1s per cycle. A reference sample was prepared and covered by 20 nm of $Al_2O_3$ using a thermal (i.e. plasma free) ALD process with TMA and $H_2O$. The reason for the difference of the $Al_2O_3$ thickness in this reference sample is that less than 20 nm $Al_2O_3$ did not provide closed layers suitable for electrical gating, which is typically observed. Raman spectroscopy was performed before and after the deposition to monitor the structural parameters like defects and strain variations.
Double-gated graphene transistors (GFET) were fabricated for measuring the electrical parameters of the graphene and the dielectrics. The process flow for device fabrication is shown in **Figure 1**a. Bottom contacts were made of Pd/Ti by e-beam evaporation on the substrate being Si covered by thermally grown $SiO_2$ (90 nm). The Si substrate acted as a global back gate. The graphene/hBN stack was then transferred and patterned by reactive ion etching. Afterwards 10 nm $Al_2O_3$ was deposited by PEALD as described above. Vias were etched by reactive ion etching to access the contact pads. Finally, the top-gate was fabricated by optical lithography, metal deposition (50 nm Al by e-beam) and lift-off. A drawing of the final device is shown in Figure 1b. All electrical measurements were performed at room temperature under ambient conditions. In addition, Raman spectroscopy was performed at the different stages of fabrication.

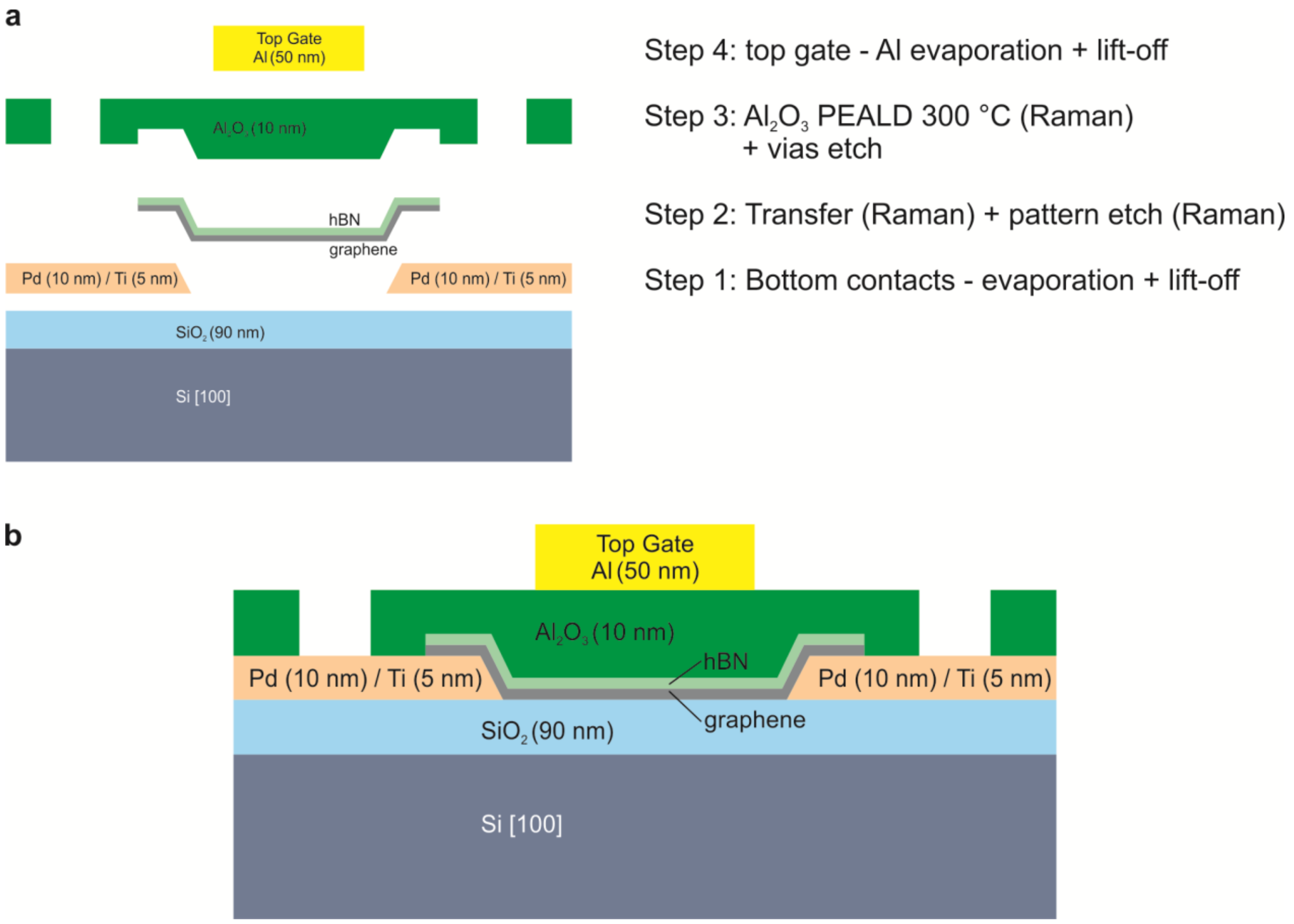

**Figure 1.** Process flow schematic for the GFET device fabrication. a) Individual building blocks and steps, b) final device.

## 2. Results and Discussion

### 2.1. Damage analysis between $Al_2O_3$ deposited by $H_2O$ and plasma process ALD

Raman spectroscopy and electrical characterization was performed to analyze the PEALD deposition of the $Al_2O_3$ on hBN/Graphene. Figure 2 summarizes the results of the Raman spectroscopy measurements for a wafer with hBN/graphene encapsulated by PEALD, a wafer with only graphene encapsulated by PEALD and a reference wafer with only graphene encapsulated by thermal ALD. Typical Raman spectra (i.e. the spectra with the median value of $I_D$ /$I_G$ for each respective map) are shown in the Figure 2a recorded directly after transfer and Figure 2b after encapsulation with $Al_2O_3$. For graphene the intensity of the D peak located at around 1350 $cm^{-1}$ provides information on the defect density of the crystal lattice.[34] After transfer, no D-peak is visible indicating high graphene quality for all three wafers. After encapsulation a clear D-peak becomes prominent for the sample encapsulated by PEALD without hBN on top of graphene, which demonstrates that the PEALD process significantly damages unprotected graphene. For the sample encapsulated by PEALD with hBN on top of graphene, the D-peak is close to the noise floor, which demonstrates an efficient protection of the graphene layer. Actually, the whole spectrum (blue curve) of this sample looks similar to the sample encapsulated by the reference process with thermal ALD (green curve). The deposition of $Al_2O_3$ also causes a decrease in the intensity ratio $I_{2D}/I_G$ and a shift in the positions of the G and 2D peaks. The $I_{2D}/I_G$ in monolayer graphene mainly depends on the doping level,[35] with higher doping leading to a lower $I_{2D}/I_G$ ratio. In addition the optical interference pattern provided by the dielectric environment also can have a significant effect on the $I_{2D}/I_G$ ratio.[36] In our samples both the doping level and the dielectric interference pattern of the substrate are changed by the deposition of $Al_2O_3$. Therefore the $I_{2D}/I_G$ ratio is not well suited parameter to monitor the doing level in the graphene. For more details about the doping level and strain in our stack before and after encapsulation, plots of the 2D-peak position vs. the G-peak position for all spectra of the Raman mappings are shown in the **Supplementary Material (S1)**.

In order to investigate the structural change due to the encapsulation in more detail, histograms of key parameters have been extracted from the mapping of the Raman spectra. Figure 2c shows the $I_D/I_G$ intensity ratio before (red) and after (green) encapsulation. Again, the sample encapsulated by PEALD without hBN shows a significant increase in the $I_D/I_G$ from 0.03 to 1.6, while the sample with hBN only shows a slight increase of the $I_D/I_G$ from 0.02 to 0.04, confirming that a single layer of hBN is efficient in protecting the graphene against plasma damage during encapsulation. The reference sample encapsulated by thermal

ALD even shows a reduction of the $I_D/I_G$ ratio. However, one needs to point out that the $I_D$ in this sample is close to the noise floor and that the peak intensities of the Raman spectra do not only depend on the crystalline structure of graphene, but also on the optical interference pattern of the substrate, which changes by the deposition of the $Al_2O_3$. The noise level is shown in the histogram of Figure 2c as solid perpendicular lines and $I_D/I_{Noise}$ is shown in the **Supplementary Information (S2)**. The reduction of $I_D/I_G$ must therefore be attributed to a reduction of the noise level due to a change in interference pattern and not to a reduction of defects. Another important parameter of the Raman spectrum of graphene is the full-width-at-half-maximum (FWHM) of the 2D peak, which mainly depends on nanoscale strain variations: A low FWHM(2D) corresponds to small strain variations,[37-39] which ultimately enables high carrier mobilities.[37-40] These variations not only depend on the quality of the graphene, but also on the smoothness of the substrate as well as the interactions of the graphene with the substrate and the encapsulation.[40-42] For example, graphene encapsulated by single crystal, multilayer hBN has a very small FWHM(2D) (less than 20 $cm^{-1}$)[41, 43] whereas encapsulation by amorphous oxides typically increases the FWHM(2D). Figure 2d shows the histograms of FWHM(2D) of the samples before and after deposition. The sample without hBN and encapsulated by PEALD shows a significant increase of the FWHM(2D) from 28 $cm^{-1}$ to 52 $cm^{-1}$. This is mainly due to the damage to the graphene caused by the plasma exposure. In contrast, the other two samples show only a moderate increase from around 29 $cm^{-1}$ to 32 (plasma ALD) or 33 $cm^{-1}$ (thermal ALD). This means that the PEALD encapsulation of the hBN/graphene sample delivers similar levels of local strain variations as the thermal ALD process. Note that a FWHM(2D) of 32 $cm^{-1}$ is state-of-the-art for CVD graphene encapsulated by a scalable method, even when compared to single crystalline CVD grown graphene flakes indicating homogeneity of the hBN monolayer transfer as otherwise holes therein would lead to local graphene damage.[44] Another parameter having an impact on the damage of the graphene is the plasma time for each cycle in the ALD process. In the **Supplementary Material (S3)** we compare the Raman spectra of $Al_2O_3$ deposited by PEALD on graphene with and without protection by hBN with short plasma time (0.1s) and with longer plasma time (2s). For the longer plasma time significant damage can be observed even with an hBN monolayer on top of the graphene, demonstrating that the deposition process also has an impact on the damage to the graphene.

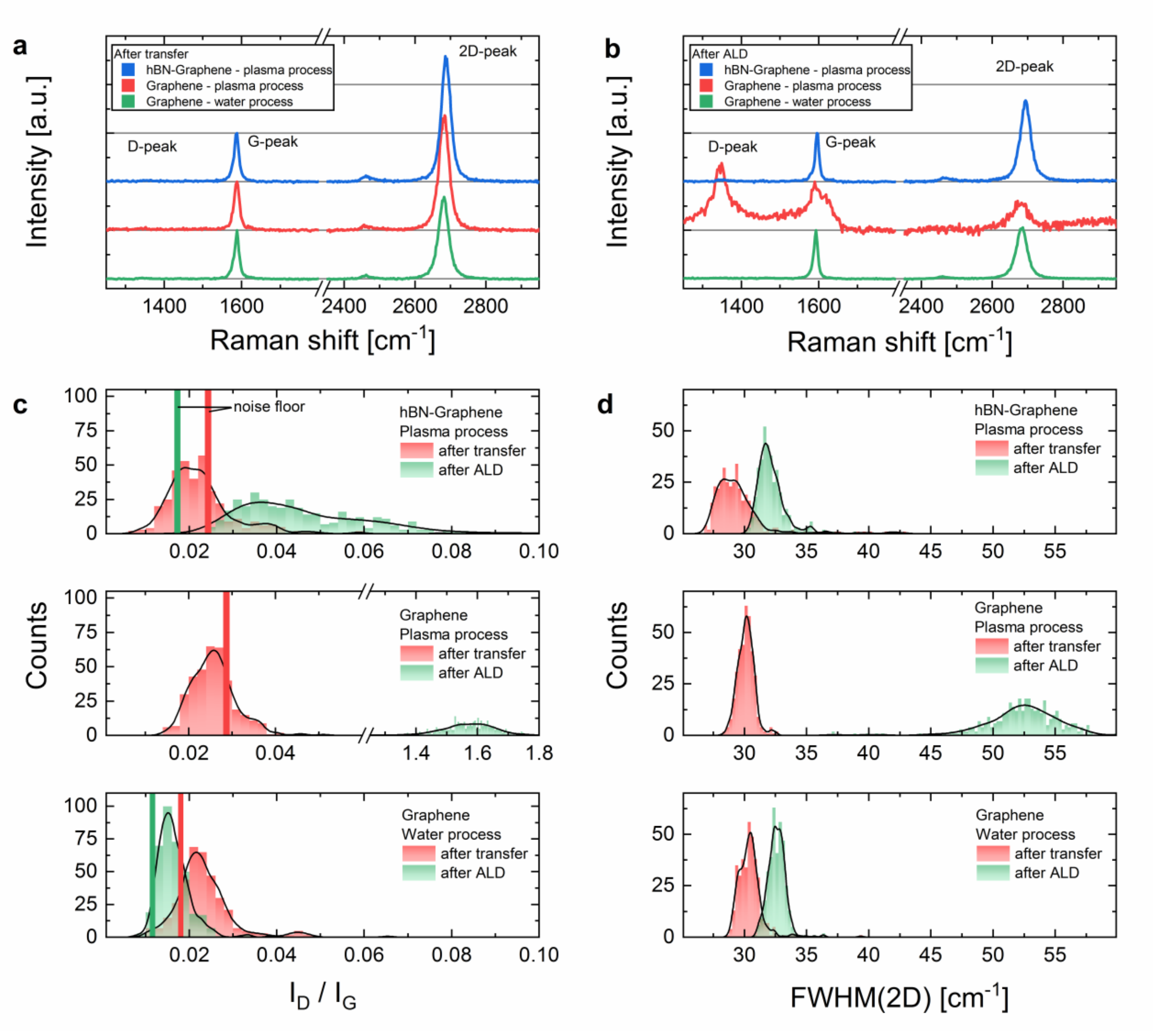

**Figure 2.** Raman analysis for $Gr/SiO_2/Si$ and $hBN/Gr/SiO_2/Si$ before and after $Al_2O_3$ deposition. a) Raman spectra for $Gr/SiO_2/Si$ and $hBN/Gr/SiO_2/Si$ after transfer. b) Raman spectra for $Gr/SiO_2/Si$ and $hBN/Gr/SiO_2/Si$ after $Al_2O_3$ deposited by PEALD and TMA-$H_2O$ ALD. c) Histogram of $I_D/I_G$ for $Gr/SiO_2/Si$ and $hBN/Gr/SiO_2/Si$ before and after $Al_2O_3$ deposition by PEALD and thermal ALD. The solid vertical lines indicate the average noise level of the acquired mappings (green before and red after encapsulation). d) Histogram of FWHM(2D) for $Gr/SiO_2/Si$ and $hBN/Gr/SiO_2/Si$ before and after $Al_2O_3$ deposition by PEALD and thermal ALD.

Double gate field effect devices were fabricated and Raman analysis was performed for all steps of the device fabrication (**Supplementary Information (S4)**). While overall the spectra are very similar and consistent with the ones shown in Figure 2, small differences are observed between the position of the D-peak and the G peak. This can be fitted as a broad peak that we attribute to amorphous carbon (a-C).[45] In our samples source of this carbon is resist residue remaining after optical lithography, which leads to a higher peak.

## 2.2 Electrical Characterization of Devices

The electrical properties of the graphene were assessed through back-gate mode and top-gate mode drain current versus gate voltage measurements (transfer characteristics) at different stages of sample fabrication. **Figure 3**a shows an optical microscope image of one double gated device. Figure 3b shows the back-gate transfer characteristic for the same device before and after encapsulation (here with grounded top-gate) measured under ambient conditions. The two-probe field effect mobility, which includes the contact resistance, is aproximately $\mu$ = 1500 cm²/Vs before and after the $Al_2O_3$ deposition, demonstrating that the PEALD process does not reduce this figure of merit. The hysteresis after the $Al_2O_3$ deposition is 2.7 V, which is quite low, given that measurements were conducted at room temperature and in ambient air, and also compared with the measurement before the $Al_2O_3$ deposition (8V, note that the sweep voltages are different before and after the $Al_2O_3$ deposition). In addition, the doping level is lower after encapsulation and is close to zero. Here, doping refers to doping by charge transfer rather than substitional doping. Similarly, low doping levels and hysteresis were observed in graphene encapsulated with $Al_2O_3$ using thermal ALD.[19] For the double gate measurements, the back-gate was kept constant and the top-gate was swept. The top-gate transfer curves for 7 different back-gates for one device can be seen in **Figure 4**a. The sweep range for the top gate voltage is from -3 to 3 V and the back-gate voltages were: 0V, +10V, -10V, +20V, -20V, +30V, -30V. The two-probe field effect mobility is aproximately $\mu$ = 500 cm²/Vs, which is lower compared to the back-gate curves. The reason for this difference is that a large area of the graphene between the contact pads and the top gate is not gated by the local top gate, which results in a large series access resistance. Subtracting this access resistance would significantly increase the extracted mobility value,[46] but as this will not lead to new insights with respect to this work, such a subtraction is not performed here. The hysteresis for the top-gate sweeps is larger compared to back-gate measurements of this device (area of the top gate), although similar to other results obtained on top-gated GFETs measured at comparable electric fields.[47] From the curves in Figure 4a we extracted the dielectric constant $\varepsilon_r$ of the $Al_2O_3$ by plotting the top gate voltage at the charge neutrality point (CNP) as a function of the applied back-gate voltage (Figure 4b). The slope of a linear fit in this plot is equal to the ratio of the capacitance of the back gate (90 nm $SiO_2$) to the top gate capacitance (10 nm $Al_2O_3$), so $\varepsilon_r$ can be calculated. The slope is 0.073, which corresponds to a top-gate equivalent oxide thickness (EOT) of 6.6 nm, assuming $\varepsilon_r$ = 3.9 for $SiO_2$. This results in a dielectric constant of 5.9 for the 10 nm $Al_2O_3$ layer. This value of $\varepsilon_r$ is similar to the ones reported in literature, e.g., for thermal ALD with TMA and $H_2O$ precursors, 4.9 – 9.1 is reported in literature depending on the deposition process and conditions of the measurements.[16, 18, 21, 48] Note that only 10 nm $Al_2O_3$ have been assumed as top-gate dielectric. The monolayer hBN and possible interfacial contamination were not included in the estimation of the dielectric constant of the $Al_2O_3$. Considering such effects, the dielectric constant of $Al_2O_3$ will become slightly larger. However, for device applications only the total capacitance or the EOT of the top-gate are important.

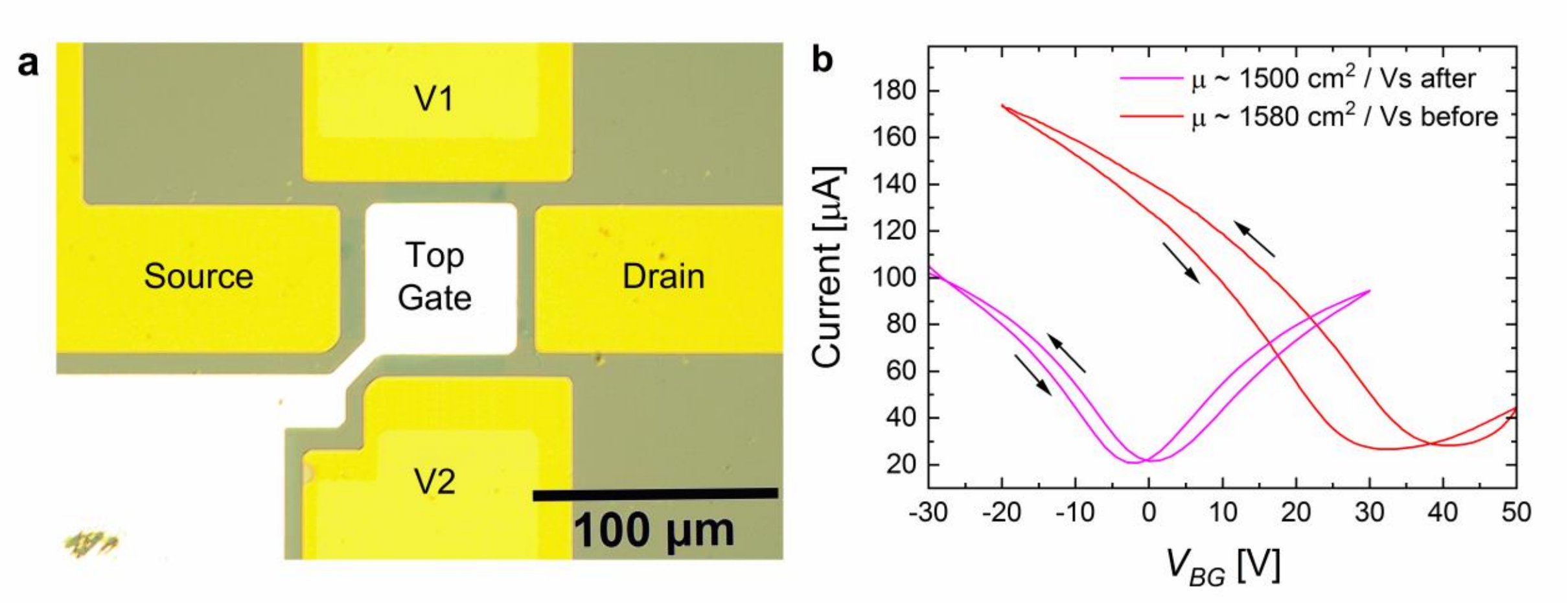

**Figure 3.** Device characterization. a) Optical microscope image of one device after top-gate fabrication. b) Transfer curves of back-gate measurements for one device before and after the $Al_2O_3$ deposition and top-gate fabrication.

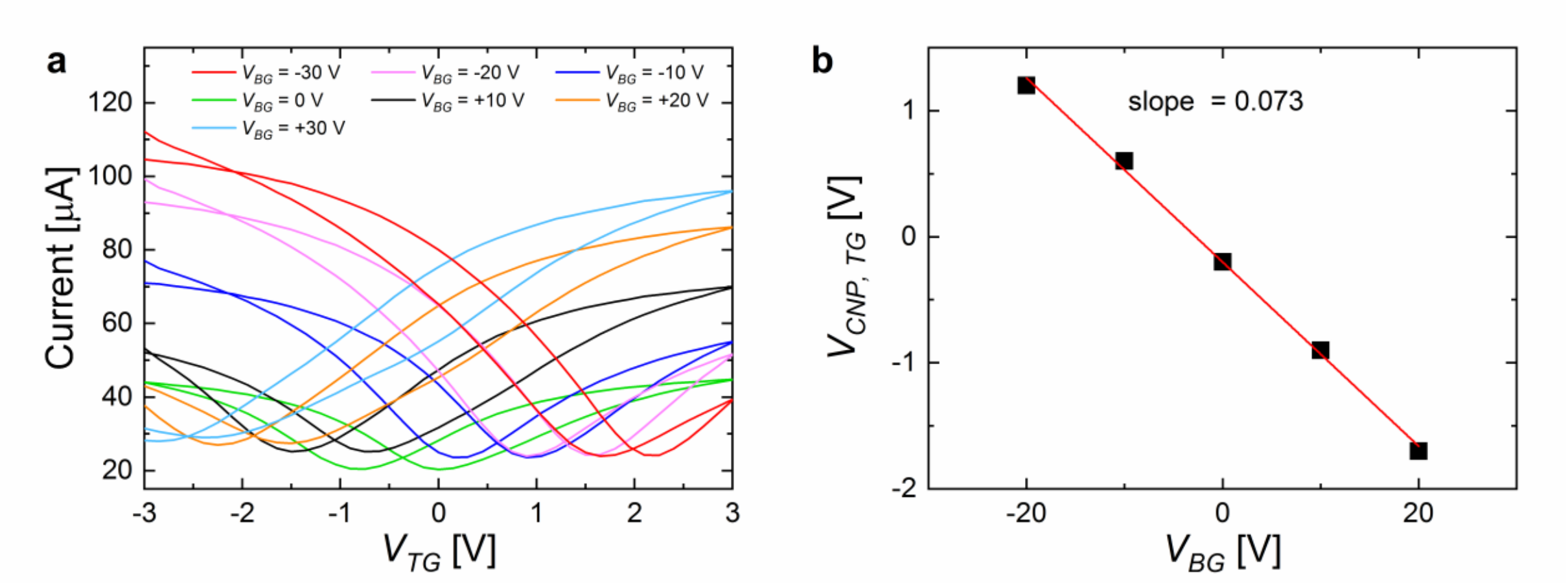

**Figure 4.** Transfer curves of top-gate measurements with the back-gate voltage kept constant for both sweeping directions. a) Sweep range: -3V to 3V, each curve (different color) was at a different constant back-gate voltage from -30V to 30V (10V steps). b) Top gate voltage at the charge neutrality point (CNP) as a function of the applied back-gate voltage.

**Figure 5** shows the top-gate leakage current for different sweeping ranges in linear (a) and log-scale (b). The gate dielectric is stable for sweeping ranges from -8 to 8 V with leakage current densities below 1 pA/µm². Above 8 V irreversible breakdown happens in this device (see inset Figure 5a). Overall, in 50 devices the breakdown voltage was measured on different wafers with values between 7.5 and 9.8 V. 8 V correspond to an electric field strength of about 8 MV/cm, which is very high for dielectrics deposited on graphene, especially keeping in mind the rather large area of the gate electrode (50x50 µm²). These values match those in reference structures without graphene for the $Al_2O_3$ ALD process.[32] Therefore, we conclude that the dielectric is not only pinhole-free, but also that the quality of the dielectric is not reduced due to the growth on top of a two-dimensional material heterostructure.

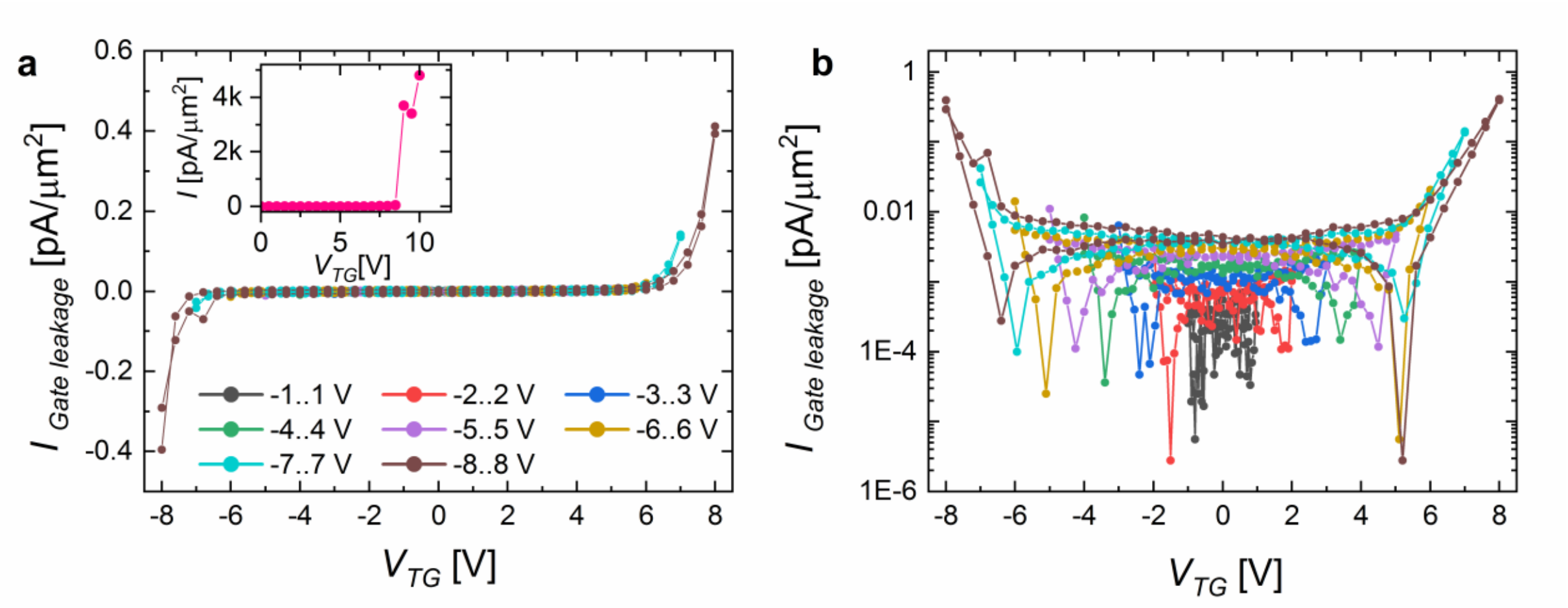

**Figure 5.** Electrical behavior of one device. a) Leakage current as a function of the gate voltage for 8 different sweep ranges from 2 V (-1 V to 1 V) to 16 V (-8V to 8V). The behavior is stable for several sweeps. Inset: Sweep range of 20V plotted from 0 to 10 V. Irreversible breakdown of the device occurs at 8 V. b) The same leakage currents of (a) plotted on a log-scale.

## 3. Conclusion

The present study investigates the deposition of $Al_2O_3$ on graphene by PEALD with and without hBN as an interfacial layer. Monolayer CVD hBN acts as an efficient protection layer against direct plasma damage, while the direct plasma process on graphene without protection caused significant damage. The dielectric properties of the PEALD $Al_2O_3$ are excellent with an EOT of 6.6 nm and a high breakdown electric field strength and gate leakage currents below 1pA/µm² at gate fields up to 8 MV/cm.

## 4. Experimental Section/Methods

*CVD of graphene and hBN*: We used CVD grown monolayer graphene and monolayer hBN from in-house sources and commercially available on 15cm Cu foils (Grolltex, San Diego, USA). The in-house growth was performed as follows: Continuous monolayer graphene films consisting of mm-sized domains were grown by CVD on Cu foils.[49] Briefly, Cu foils (99.9 %) were oxidized in air at 250 °C for 5 minutes, followed by annealing in Ar (50 mbar, 30 mins), and subsequently by $H_2$ annealing (12 mbar $H_2$, 38 mbar Ar, 60 mins) at around 1075 °C. Graphene was grown by introducing a $CH_4/H_2/Ar$ mixture ($1.2\times10^{-2}$ mbar, 9 mbar and 41 mbar respectively) for 6 hours. Monolayer hBN was grown by CVD on Fe foils as described in prior work.[49-51] Briefly, 0.1 mm Fe foils (99.8% purity) were oxidized in air at 350 °C for 5 mins, followed by annealing in $1\times10^{-2}$ mbar Ar at 980 °C for 20 mins to remove localized impurities in the foil and create a minute oxygen reservoir. A brief reducing treatment of $3\times10^{-3}$ mbar acetylene was applied for 5 min and a 20 mins dose of ammonia at $1\times10^{-2}$ mbar. Borazine was added together with ammonia with partial pressure of $8\times10^{-4}$ mbar for 90 mins to grow a continuous monolayer hBN film consisting of domains of hundreds of microns.

*Wet transfer using PMMA and Cu etching*: Some experiments were conducted with monolayer graphene and monolayer hBN grown by CVD on metal foils commercially available on 15 cm Cu foils (Grolltex, San Diego, USA). The hBN/graphene stack was prepared by a sequential wet chemical transfer of the layers from the growth substrate (Cu) using PMMA as supporting layer.[52] First, the hBN was released by etching the copper foil, rinsed in water and transferred to graphene on copper foil. Subsequently, the hBN/graphene stack (between 1x1 and 4x4 $cm^2$) was released by again etching the copper foil and, after rinsing in water, transferred to a 6-inch Si wafer covered by 90 nm thermally grown $SiO_2$. In the reference samples only graphene was transferred without hBN.

*Wet transfer using PC and electrochemical delamination*: For the graphene and hBN that were grown in-house, the following transfer method was used: After controlled humidity oxidation of the hBN/Fe interface, and coating with a support polymer (poly-bisphenol-A-carbonate, PC), the PC/hBN stack was detached via electrochemical delamination.[53] This was directly transferred to graphene/Cu and after drying and curing (160 °C, 5 mins), the PC/hBN/graphene structure was detached from Cu via electrochemical delamination. Finally, the stack was transferred onto the target pre-patterned wafer and PC was dissolved in chloroform. This material and transfer were just used for device fabrication.

*$Al_2O_3$ ALD deposition*: After the transfer, the samples were encapsulated by PEALD with 10 nm of $Al_2O_3$ in the Oxford Instruments Atomfab™ ALD and FlexAL™ system using TMA and oxygen plasma as precursors. For the Atomfab™ ALD system the deposition temperature was 300°C, the plasma time was 0.1s and the total cycle time was 0.9s. Reference samples were prepared and covered by 20 nm of $Al_2O_3$ using a thermal (i.e. plasma free) ALD process with TMA and $H_2O$ (10s cycle time) in a FlexAL™ ALD at 150°C.

*Raman characterization*: A Horiba XPlora Raman Spectrometer with a 100X microscope objective was used for Raman measurements. The excitation wavelength was 532 nm. Raman spectroscopy was performed before and after the deposition to monitor the structural parameters like defects and strain variation. 10 x 10 $\mu m^2$ areas with 121 spectra each were measured at several locations for each sample. The peaks of each spectrum were fitted by a Lorentzian function. From the fits the ratio between the D and G peak intensities as well as the full width-half maximum (FWHM) of the 2D peak was determined was determined and histograms were created for each sample.

*Device Fabrication*: Double-gated graphene transistors (GFET) were fabricated according to the work flow depicted in figure 1a. Bottom contacts were made of Pd/Ti by e-beam evaporation on the substrate, i.e. Si covered by thermally grown $SiO_2$ (90 nm). The Si substrate acted as a global back gate. Afterwards the graphene/hBN stack was transferred (as described above) and patterned by reactive ion etching. Afterwards 10 nm $Al_2O_3$ was deposited by PEALD in an Oxford Instruments Atomfab™ ALD system as described above.

Vias were etched by reactive ion etching to access the contact pads. Finally, the top-gate was fabricated by optical lithography, metal deposition (50 nm Al by e-beam evaporation) and lift-off. A drawing of a final device is shown in the Figure 1(b). It has been confirmed in previous works using Raman spectroscopy, XPS and TEM that these films are monolayer films of hexagonal boron nitride. [50, 51, 54]

*Electrical Measurements*: The electrical properties of the graphene film and of the dielectric was measured in double-gated graphene transistors, having a global back-gate and a local top-gate electrode. The back-gate transfer curves were performed to compare the mobility before and after encapsulation. The breakdown voltage and leakage current were extracted from the top-gate measurements and the dielectric constant was extracted from double-gate measurements. The Si substrate acted as global back gate. All electrical measurements were performed at room temperature under ambient conditions.

**Supporting Information**

Supporting Information is available from the Wiley Online Library or from the author.

**Acknowledgements**

This work has received funding from the European Union’s Horizon 2020 research and innovation programme under grant agreements ULISSES (825272), Graphene Flagship Core 2 and 3 (785219, 881603), 2D-EPL (952792) and ECOMAT (796388), and from the German Ministry of Education and Research BMBF within the project GIMMIK (03XP0210).

Received:

Revised:

Published online:

*Bárbara Canto, Martin Otto, Michael J. Powell, Vitaliy Babenko, Aileen O'Mahony, Harm Knoops, Ravi S. Sundaram, Stephan Hofmann, Max C. Lemme and Daniel Neumaier**

**Deposition of thin high-quality $Al_2O_3$ films on graphene using plasma-enhanced atomic layer deposition**

The deposition of $Al_2O_3$ by plasma-enhanced ALD on graphene protected by a monolayer of hBN is investigated. Monolayer hBN acts as an efficient protection layer against direct plasma damage of graphene as shown by Raman spectroscopy. Top-gated devices (GFETs) fabricated with thin dielectric oxides on hBN/graphene exhibit consistently low leakage currents below 1pA/µm² and high breakdown voltages.

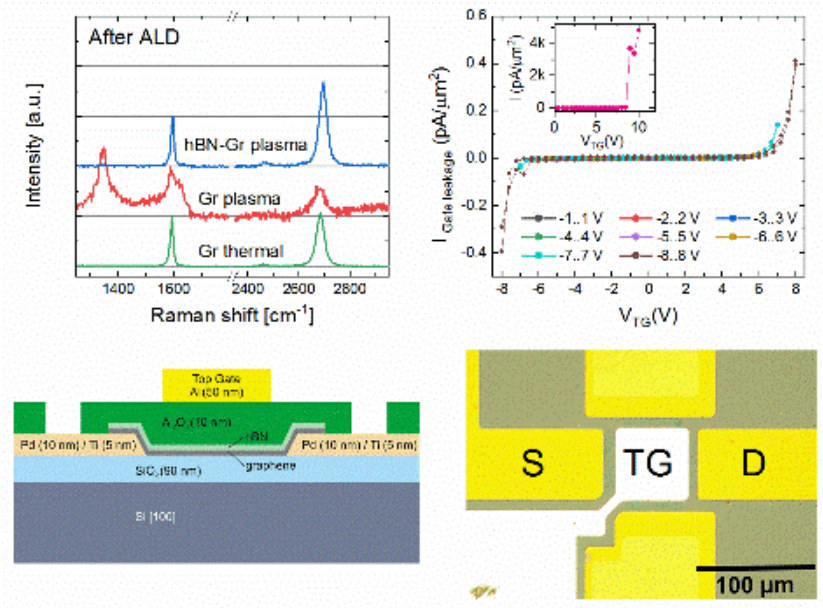

# Supporting Information

## Deposition of thin high-quality $Al_2O_3$ films on graphene using plasma-enhanced atomic layer deposition

*Bárbara Canto, Martin Otto, Michael J. Powell, Vitaliy Babenko, Aileen O'Mahony, Harm Knoops, Ravi S. Sundaram, Stephan Hofmann, Max C. Lemme and Daniel Neumaier*

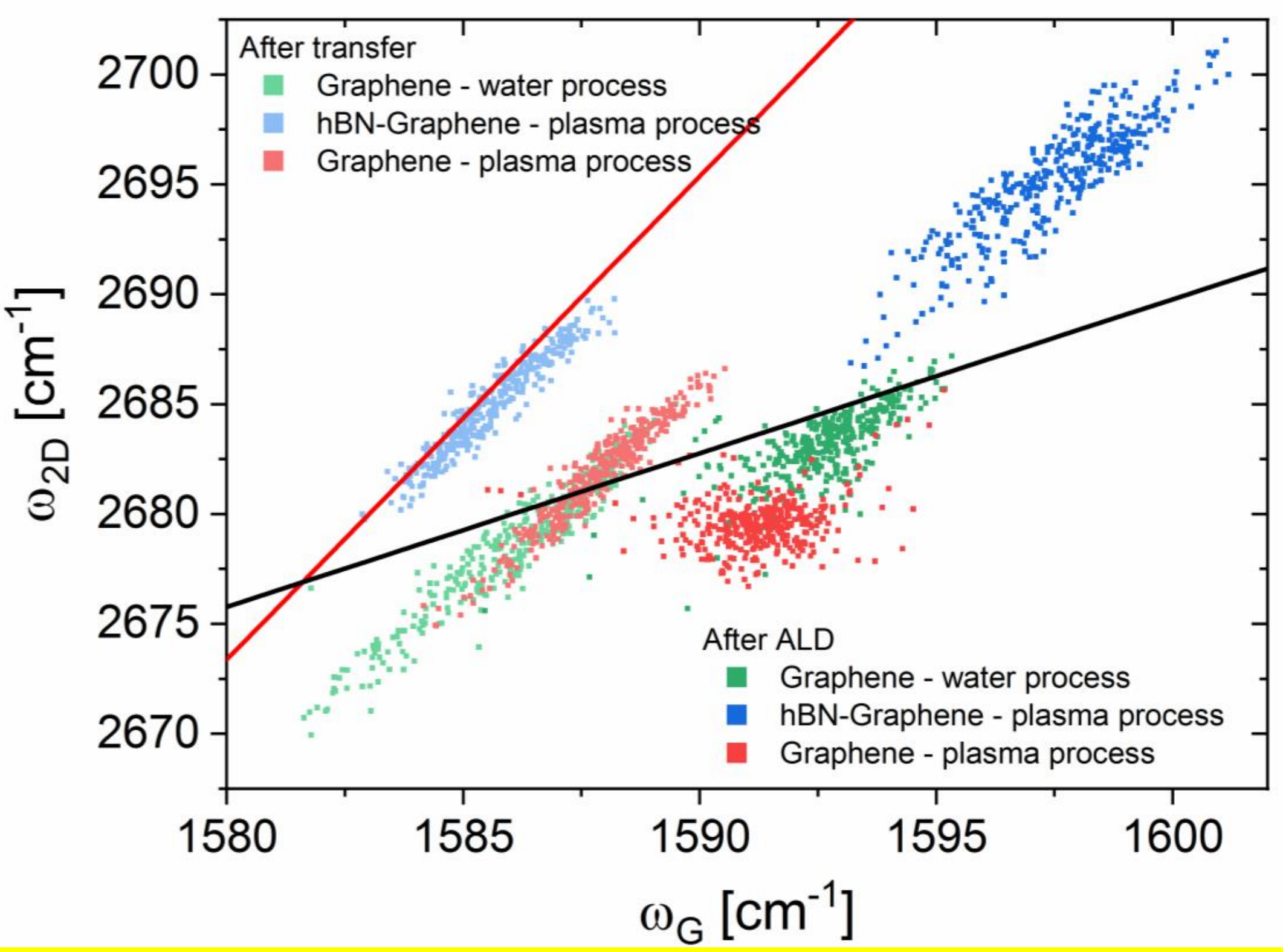

**Figure S1.** Scatter plot of the 2D peak position as a function of the position of the G-peak. The red line denotes direction of shifts of the peak positions due to strain and the black line represents the shifts due to doping of the graphene.[53] It can be seen that there is compressive strain in the hBN/Graphene stack which remains unchanged after oxide deposition by ALD. The deposition increases the doping level for all films regardless of ALD method (plasma or thermal) and for both graphene and hBN/graphene. Interestingly, the doping after deposition is almost the same for hBN/graphene and graphene (thermal ALD) even though the hBN/graphene had lower doping initially. The damage done to the unprotected graphene by PEALD is substantial, which also strongly affects the peak positions, making it difficult to interpret the plot in terms on strain and doping for this particular sample.

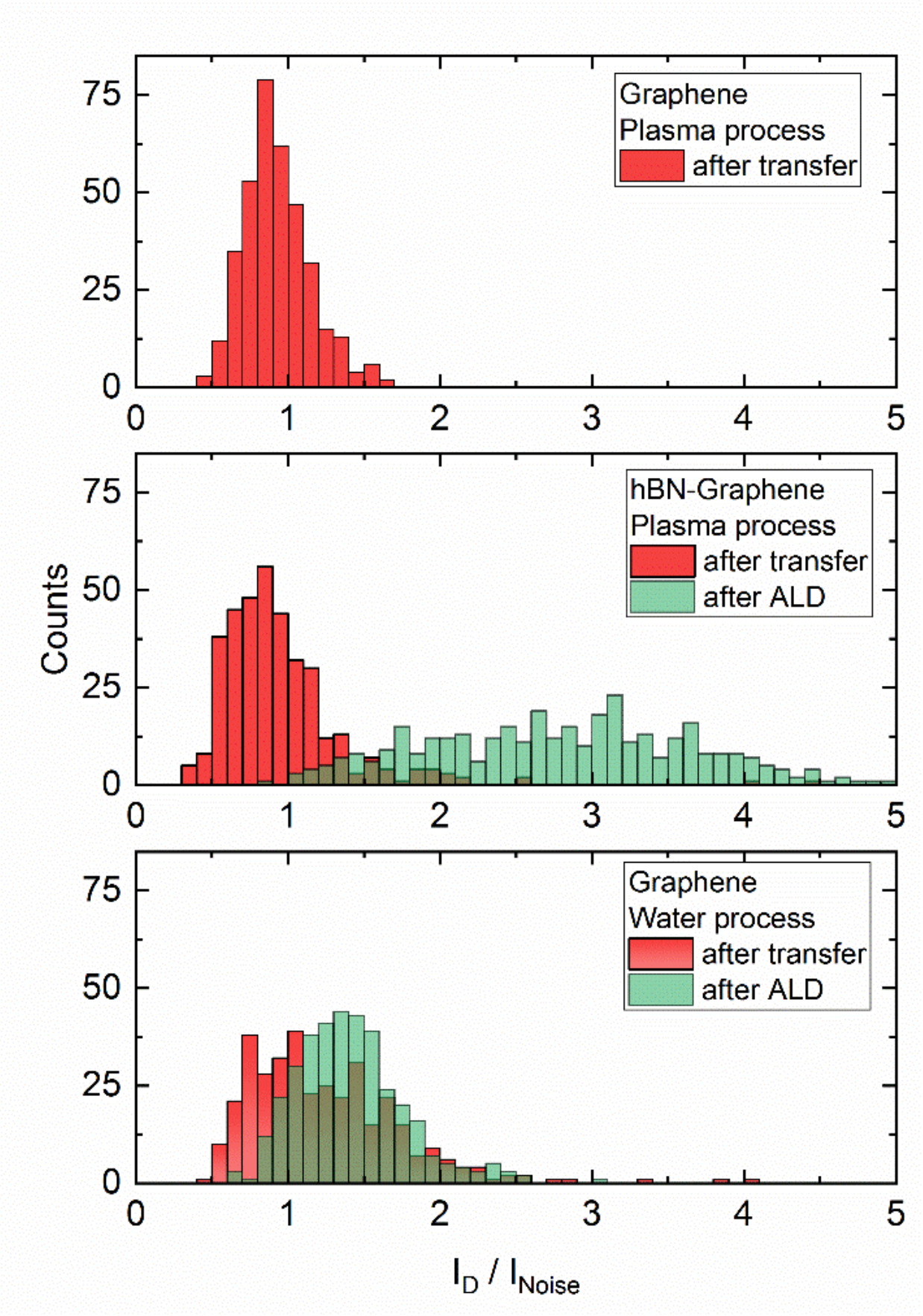

**Figure S2.** Raman noise analysis of three different samples comparing the intensity of the D-peak fit to the maximum intensity of the noise for each spectrum. For values smaller than 1, the D-peak is essentially indistinguishable from the noise.

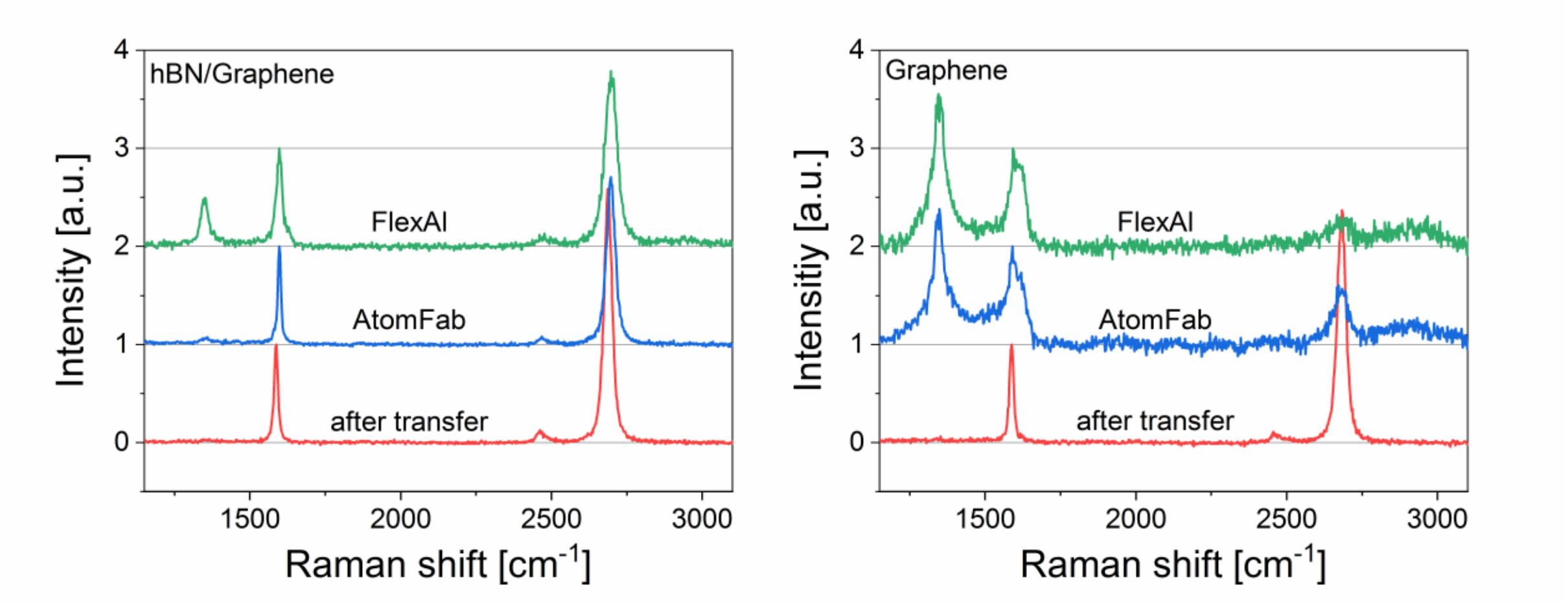

**Figure S3.** Raman spectra comparing AtomFab™ and FlexAL™ $Al_2O_3$ depositions on hBN/graphene (left) and graphene only (right). The depositions were performed by PEALD at 300 °C in both machines. The main difference between the processes are plasma time and plasma power: 0.1 s and 100 W for the AtomFab™ system and 2 s and 400 W for the FlexAl™.

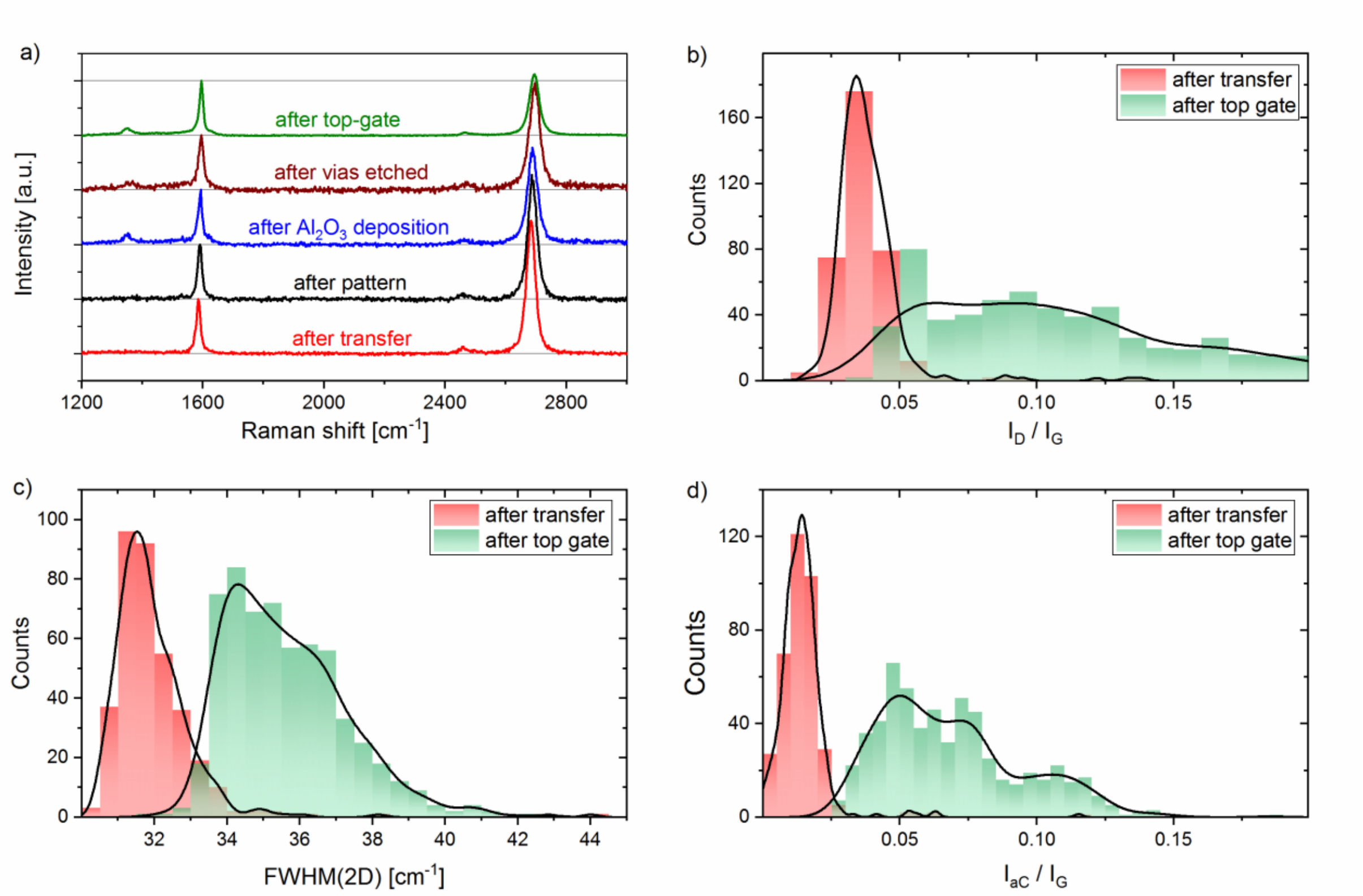

**Figure S4.** Raman analysis of a double gate GFET. a) Raman spectra for all steps of fabrication. b) Histogram of $I_D/I_G$, c) Histogram of FWHM(2D), d) Histogram of $I_{\alpha\text{-}C}/I_G$ for the first (after transfer) and the last (after top gate fabrication) steps of fabrication (involving graphene) showing the small increase of damage after the $Al_2O_3$ deposition.